\documentclass[12pt]{article}

\usepackage{amsmath,amssymb,amsthm,epsf,epsfig}
\usepackage{amsmath,amssymb}

\newcommand{\ed}{\end{document}}
\newcommand{\be}{\begin{equation}}
\newcommand{\ee}{\end{equation}}
\newcommand{\pa}{\partial}

\begin{document}
\begin{center}
\large{\textbf{Gauge Invariant Extension of Linearized Horava Gravity}}\\
\end{center}
\begin{center}
Sudipta Das {\footnote{E-mail: sudipta.das\_r@isical.ac.in} and
Subir Ghosh {\footnote{E-mail: sghosh@isical.ac.in}\\
Physics and Applied Mathematics Unit, Indian Statistical
Institute\\ 203 B.T.Road, Kolkata 700108, India}}
\end{center}

\vskip 2.5cm

\begin{center}
{\it{\textbf{Abstract}}}\\
\end{center}

In the present paper we have constructed a gauge invariant extension of a generic Horava Gravity (HG) model
(with quadratic curvature terms) in linearized version in a systematic procedure. No additional fields are
introduced. The linearized HG model is explicitly shown to be a gauge fixed version of the Einstein Gravity (EG)
thus proving the Bellorin-Restuccia conjecture in a robust way. In the process we have explicitly computed the correct
Hamiltonian dynamics using Dirac Brackets appearing from the Second Class Constraints present in the HG model. We comment
on applying this scheme to the full non-linear HG.

\newpage

{\it{Introduction and summary of our work}}: It is well known that in a generic
quantum field theory higher derivative terms improve its ultraviolet behavior
but, unfortunately, in the context of Einstein Gravity (EG) the above generates
ghosts \cite{stelle} thus rendering the covariant higher derivative extension
unacceptable. To lift this impasse Horava \cite{horava} has proposed an
ingenious idea of introducing higher derivative terms in the spatial sector
only without modifying the kinetic part. The advantage of better ultraviolet
behavior together with the non-appearance of ghosts in the Horava Gravity (HG)
\cite{horava}
however is achieved at a steep price: the full diffeomorphism invariance is
replaced by  foliation-preserving diffeomorphism invariance only. According to
the works of \cite{blas1, blas2, kob, henn, li}, this loss of symmetry makes HG
inconsistent as it induces a peculiar constraint structure and an extra
dynamical mode of a non-canonical nature,
(besides the graviton), and HG fails to match EG in low energy regime, a
prerequisite of any viable extension of EG. Very recently the constraint
structure of HG has been re-examined in
\cite{bellorin1, bellorin2} and it is shown that, contrary to previous claims in
\cite{blas1, blas2, kob, henn, li}, HG is a consistent theory but
the presence of extra mode is unavoidable in the full theory. The role of the potential function has been
studied in \cite{don}.

In the present work we study linearized HG
\be
S= \int dt d^3 x \sqrt{g} N (K_{ij} K^{ij} - \lambda K^2
+ A R + B R_{ij} R^{ij} + C R^2 ).
\label{faction}
\ee
and explicitly demonstrate the following:\\
(i) HG is a completely consistent constraint system having a conventional form
of Second Class Constraints. \\
(ii) The $\lambda, A$ and $C$ terms do not play any role in linearized HG. \\
The above observations corroborate with \cite{bellorin1, bellorin2}.\\
(iii) {\it{Most importantly we provide a systematic way of further extending HG to a
gauge invariant theory}}. Interestingly enough,
we recover the linearized EG modified by the $B$-term (\ref{faction})
contribution only. This form of improvement has been
suggested in a heuristic way in \cite{blas1} for linearized HG. We follow a
general scheme developed in \cite{mitra, vyth}
which can be applied to the full HG as well. The latter work is presently under
study.

{\it{Linearized Horava Gravity}}: We start with the following form of HG
(\ref{faction}),
\be
S=\int dt {\cal{L}}=\int dt d^3 x \sqrt{g} N (G^{ijkl}
K_{ij} K_{kl} + A R + B R_{ij} R^{ij} + C R^2 ) $$$$
= \int dt d^3 x \sqrt{g} N (K_{ij} K^{ij} - \lambda K^2
+ A R + B R_{ij} R^{ij} + C R^2 ).
\label{fullaction}
\ee

Here $g_{ij}$ is the spatial metric, $A, B, C$ are
three dimension-full parameters of the theory,
$N$ is the Lapse function, $R$ is the spatial Ricci
scalar and $K_{ij}$ is the extrinsic
curvature defined as
\be
K_{ij}=\frac{1}{2 N}(\pa_0 g_{ij} - \nabla_i N_j - \nabla_j N_i).
\label{k}
\ee
with $N_i(x,t)$ is the
Shift vector in ADM formalism  \cite{adm} and the generalized De Witt metric
$G^{ijkl}$ is defined as
\be
G^{ijkl} = \frac{1}{2} (g^{ik}g^{jl} + g^{il}g^{jk}) - \lambda g^{ij} g^{kl}.
\label{dewitt}
\ee
For $\lambda = 1$, $A=1$ and $B=C=0$ HG reduces to EG.

Consider the following perturbations to the metric:
\be
g_{ij}=\delta_{ij}+h_{ij}~~~,~~~N=1+n~~~,~~~N_i=n_i. \label{per}
\ee

Under these perturbations (\ref{per}), the expressions for the
extrinsic curvature and the Ricci curvature turn out to be
\be
K_{ij}=\frac{1}{2}(\pa_0 h_{ij} - \pa_i n_j - \pa_j n_i)~~,~~
K=\delta^{ij}K_{ij}=\frac{1}{2}(\pa_0 h - 2 \pa_i n^i), $$$$
R_{ij}=\frac{1}{2}(\pa^k \pa_i h_{jk}+\pa^k \pa_j h_{ik}-\pa^2 h_{ij} -
\pa_i \pa_j h)~~,~~
R=\pa_i \pa_j h^{ij} - \pa^2 h. \label{kr}
\ee

Using the above expressions (\ref{kr}) and the relation
\be
\sqrt{g} R=\frac{1}{2} h_{ij}\left(-R^{ij}+\frac{1}{2}
\delta^{ij} R\right)
\label{gr}
\ee
in the action (\ref{fullaction}) we obtain the
Lagrangian density ${\cal{L}}$ of second order in $h$:
\be
{\cal{L}}=\frac{1}{4}[\pa_0 h_{ij} \pa_0 h^{ij} - \lambda(\pa_0 h)^2
-4(\pa_0 n_i)(\pa_j h^{ij} - \lambda \pa^i h)
+ (4 \lambda -2)n_i (\pa^i \pa^j n_j) - 2 n_i \pa^2 n^i] $$$$
+ \frac{1}{4} h_{ij} (\pa^2 h^{ij} - 2 \pa_k \pa^i h^{jk}
+ 2 \pa^i \pa^j h - \delta^{ij} \pa^2 h)
+ A n (\pa_i \pa_j h^{ij} - \pa^2 h)
+ C (\pa_i \pa_j h^{ij} - \pa^2 h)(\pa_k \pa_l h^{kl} - \pa^2 h)
$$$$ + \frac{B}{4} (\pa^k \pa_i h_{jk} - \pa^2 h_{ij}
+ \pa^k \pa_j h_{ik} - \pa_i \pa_j h)
(\pa_l \pa^i h^{jl} - \pa^2 h^{ij}
+ \pa_l \pa^j h^{il} - \pa^i \pa^j h). \label{lag}
\ee

In the above expression (\ref{lag}), we used the following notation:
$$ h = \delta^{ij} h_{ij}~~~~,~~~~\pa^2 = \pa_i \pa^i = \delta^{ij}
\pa_i \pa_j.$$

{\it{Hamiltonian of linearized Horava Gravity}}: From this Lagrangian
(\ref{lag}) we
obtain the conjugate momenta as
\be
p \equiv \frac{\pa {\cal{L}}}{\pa (\pa_0 n)} = 0~~~,~~~
p^i \equiv \frac{\pa {\cal{L}}}{\pa (\pa_0 n_i)} =
-(\pa_j h^{ij} - \lambda \pa^i h), \label{p}
\ee
\be
\pi^{ij} \equiv \frac{\pa {\cal{L}}}{\pa (\pa_0 h_{ij})} =
\frac{1}{2} (\pa_0 h^{ij} - \delta^{ij} \lambda (\pa_0 h)).
\label{pi}
\ee

Taking the trace of the relation (\ref{pi}), we can write
$\pa_0 h_{ij}$ in terms of $\pi_{ij}$ as
\be
\pa_0 h^{ij} = 2 \left(\pi^{ij}+\frac{\lambda}{1-3 \lambda}
\delta^{ij} \pi \right). \label{delh}
\ee

Using the relations (\ref{p}), (\ref{pi}), (\ref{delh}) in (\ref{lag})
we get the Hamiltonian density,
\be
{\cal{H}} = p^i (\pa_0 n_i) + \pi^{ij} (\pa_0 h_{ij}) - {\cal{L}} $$$$
= \pi_{ij} \pi^{ij} - \frac{\lambda}{3 \lambda - 1} \pi^2 -
\frac{1}{2} (2 \lambda - 1) n_i (\pa^i \pa^j n_j) + \frac{1}{2}
n_i \pa^2 n^i $$$$ - \frac{1}{4} h_{ij} (\pa^2 h^{ij} - 2 \pa_k \pa^i h^{jk}
+ 2 \pa^i \pa^j h - \delta^{ij} \pa^2 h)
- A n (\pa_i \pa_j h^{ij} - \pa^2 h)
- C (\pa_i \pa_j h^{ij} - \pa^2 h)(\pa_k \pa_l h^{kl} - \pa^2 h)
$$$$ - \frac{B}{4} (\pa^k \pa_i h_{jk} - \pa^2 h_{ij}
+ \pa^k \pa_j h_{ik} - \pa_i \pa_j h)
(\pa_l \pa^i h^{jl} - \pa^2 h^{ij}
+ \pa_l \pa^j h^{il} - \pa^i \pa^j h). \label{ham1}
\ee

{\it{Constraint analysis a la Dirac}}: We now perform a standard Hamiltonian
constraint analysis using the Dirac formalism \cite{dirac}, which we discuss
below
in brief.\\
From a given Lagrangian, one starts by computing the conjugate
momentum $p=\frac{\partial L}{\partial \dot q}$ of a generic
variable $q$ and identifies the relations that do not contain
time derivatives as (Hamiltonian) constraints. A constraint
is classified as First Class Constraint (FCC)
when it commutes with all other constraints and the set
of constraints which do not commute are called Second
Class Constraints (SCC). First Class constraints
generate gauge invariance. For systems containing
Second Class constraints, one has
to replace the Poisson brackets by Dirac brackets to properly
incorporate the Second Class constraints. If $(\{\psi_{\rho}^{i},
\psi_{\sigma}^{j}\}^{-1})$ is the $(i j)$-th element of the inverse
constraint matrix where $\psi^{i}(q,p)$ is a set of Second Class
constraints, then the Dirac bracket between two generic variables
$\{A(q,p), B(q,p)\}_{DB}$ is given by
\be \{A, B\}_{DB} = \{A, B\} - \{A,
\psi_{\rho}^{i}\} (\{\psi_{\rho}^{i}, \psi_{\sigma}^{j}\}^{-1})
\{\psi_{\sigma}^{j}, B\}, \label{ddb}
\ee
where $\{~~,~~\}$ denotes Poisson brackets. In this framework,
the constraints $\psi_{\rho}^{i}$ are ``strongly'' zero since
they commute with any generic variable $A$: $\{A,
\psi_{\rho}^{i}\}_{DB} = \{\psi_{\rho}^{i}, A\}_{DB}=0$.

From the Lagrangian (\ref{lag}), we get four primary constraints
\be
\psi \equiv p, \label{consp}
\ee
\be
\phi^i \equiv p^i + \pa_j h^{ij} - \lambda \pa^i h. \label{consphi}
\ee

Requiring the time persistence of the above four constraints (\ref{consp}),
(\ref{consphi}) we get the following four secondary constraints
\be
\chi_1 \equiv {\dot{\psi}} = \left\{\psi, \int
{\cal{H}}(y) d^3 y \right\} = \pa_i \pa_j h^{ij} - \pa^2 h, \label{chi1}
\ee
\be
\eta^i \equiv {\dot{\phi}}^i = \pa_j \pi^{ij} + \frac{1}{2}
(2 \lambda -1) \pa^i \pa^j n_j - \frac{1}{2} \pa^2 n^i. \label{eta1}
\ee

Time persistence of the constraint (\ref{eta1})
is trivially satisfied, i.e. $\dot{\eta^i} = 0$. However, $\dot \chi _1$ yields
a tertiary constraint
\be
\chi_2 \equiv {\dot{\chi_1}} = \pa_i \pa_j \pi^{ij} -
\frac{\lambda-1}{3 \lambda -1} \pa^2 \pi. \label{chi}
\ee

The chain stops since the constraints $\chi_1, \chi_2 $ (in  (\ref{chi1}) and
(\ref{chi})
respectively) constitute a pair of SCC. The constraints $\psi, \phi ^i, \eta ^j$
(in (\ref{consp}), (\ref{consphi}) and (\ref{eta1}) respectively)
are FCC. {\it{Clearly we do not find any abnormality in the constraint structure
as claimed earlier in}} \cite{blas1, blas2, kob, henn, li}.
A quick count of the dynamical degrees of freedom
\be
No.~ of~dynamical ~d.o.f~=(total~ No.~of ~d.o.f~)-(2 \times~ No.~of
~FCC)~-(No.~of
~SCC)$$$$
=20-(2 \times 7)-2=4,
\ee
shows that the system has $4$ d.o.f. in phase space that correctly represent the
graviton. Regarding the comment in \cite{li} about the viability of
perturbative analysis in obtaining the number of d.o.f. we believe our results
are indeed valid in the present weak gravity regime.

Let us choose the gauge $n_i=0$, (as is customary). This does not affect the
bracket structure of the remaining
d.o.f. The constraints (\ref{eta1}) simplifies to
\be
\eta^i \equiv \pa_j \pi^{ij}. \label{etafinal}
\ee
Using the constraints (\ref{etafinal}) the tertiary constraint (\ref{chi})
becomes
\be
\chi_2 \equiv \pi, \label{chi2}
\ee
provided $\lambda \neq 1$.

Using the constraints (\ref{chi1}), (\ref{chi2}) and the gauge $n_i=0$, the
Hamiltonian (\ref{ham1}) now reduces to
\be
{\cal{H}} = \pi_{ij} \pi^{ij} - \frac{1}{4} h_{ij} (\pa^2 h^{ij} - 2 \pa_k \pa^i
h^{jk}
+ 2 \pa^i \pa^j h - \delta^{ij} \pa^2 h)
$$$$ - \frac{B}{4} (\pa^k \pa_i h_{jk} - \pa^2 h_{ij}
+ \pa^k \pa_j h_{ik} - \pa_i \pa_j h)
(\pa_l \pa^i h^{jl} - \pa^2 h^{ij}
+ \pa_l \pa^j h^{il} - \pa^i \pa^j h). \label{hamiltonian}
\ee
Let us pause to note the following point: As we had advertised in the first
section, the parameters $ A, C$ are absent in the weak limit
but the role of $\lambda $ is interesting. Apparently $\lambda $ has disappeared
from consideration but it still has a a non-trivial
effect in an indirect way: the present constraint structure of HG having both
FCC and SCC
is distinct from EG constraint structure (the latter having
only FCC). As mentioned below (\ref{chi}), in HG we consider $\lambda \neq 1$
otherwise for $\lambda =1$ the constraint structure collapses
to only FCCs that is true for EG.

{\it{Dirac Brackets and Hamiltonian dynamics}}: For the two SCCs, $\chi_1$ and
$\chi_2$ ((\ref{chi1}), (\ref{chi2}) respectively), the constraint matrix and
its
inverse are given below:
\begin{equation}
\{\chi_1(x)~,~\chi_2(y) \}=
 \left (
\begin{array}{cc}
 0 &  -2 \pa^2 \delta^3 (x-y)\\
2 \pa^2 \delta^3 (x-y) &  0
\end{array}
\right ) $$$$ \{\chi_1(x)~,~\chi_2(y)\}^{-1}=
 \left (
\begin{array}{cc}
 0 &  \frac{1}{2 \pa^2} \delta^3 (x-y)\\
-\frac{1}{2 \pa^2} \delta^3 (x-y) &  0
\end{array}
\right ). \label{mat}
\end{equation}

Using (\ref{mat}) in
the definition of the Dirac brackets (\ref{ddb}), we have
\be
\{ h_{ij}(x), \pi^{kl}(y) \}_{DB} =
\frac{1}{2} \left(\delta^k_i \delta^l_j + \delta^k_j \delta^l_i - \delta_{ij}
\delta^{kl}
+ \delta_{ij} \frac{\pa^k \pa^l}{\pa^2}\right) \delta^3 (x-y), $$$$
\{ h_{ij}(x), h_{kl}(y) \}_{DB} =\{ \pi^{ij}(x), \pi^{kl}(y) \}_{DB} =0.
\label{hpi}
\ee
Clearly the mixed bracket has a non-canonical structure.

It is straightforward to exploit the Dirac bracket (\ref{hpi}) to compute the
Hamiltonian equations of motion:
\be
{\dot{h}}_{ij} = \left\{h_{ij} , \int d^3 y {\cal{H}}(y)\right\}_{DB} =
2 \pi_{ij} + \delta_{ij}
\frac{\pa^k \pa^l}{\pa^2} \pi_{kl}. \label{doth}
\ee
\be
{\dot{\pi}}_{ij} = \left\{\pi_{ij} , \int d^3 y {\cal{H}}(y)\right\}_{DB}
$$$$
= \frac{1}{2}\left[\pa^2 h_{ij}-\pa^k \pa_i h_{jk} - \pa^k \pa_j h_{ik}+\pa_i
\pa_j h
+B \left(\pa^2 (\pa^2 h_{ij}) - \pa^2 \pa^k \pa_i h_{jk} - \pa^2 \pa^k \pa_j
h_{ik}
+\pa^2 \pa_i \pa_j h \right) \right]. \label{dotpi}
\ee

Finally we recover the equation of motion for $h_{ij}$,
\be
\Box h_{ij} = \pa^k \pa_i h_{jk}
+ \pa^k \pa_j h_{ik} - \pa_i \pa_j h
+ B(\pa^2 \pa^k \pa_i h_{jk} + \pa^2 \pa^k \pa_j h_{ik}
- \pa^2 \pa_i \pa_j h - (\pa^2)^2 h_{ij}), \label{em}
\ee
where $$\Box h_{ij} = -{\ddot{h}}_{ij} + \pa^2 h_{ij}.$$

Taking the trace of the above equation (\ref{em}) and imposing
the constraint (\ref{chi1}) we have
\be
\ddot{h} = 0, \label{emh}
\ee
which clearly shows that the scalar $h$ does not propagate at all in case
of the HG model.

Hence, together with (\ref{emh}), (\ref{em}) correctly represents the dynamics
of a spin-2 field $h_{ij}$, modified by the higher derivative terms.
In case of EG theory, the dynamics
of the spin-2 field is also governed by the same equation (\ref{em}),
hence non-propagation of the scalar $h$ holds true for EG theory. This
constitutes first part
of our work that is proving the consistency of the (linearized) HG.
Our result is consistent
with \cite{kim, park}.

{\it{Gauge invariant extension of linearized HG}}: Gauge invariant theories are
ubiquitus in modern quantum field theoretic framework.
Gauge invariance plays an essential role in the quantization programme. In the
prsent context of HG, the weaknesses of the theory principally
result from a loss of gauge invariance, (i.e. loss of full diffeomorphism
invariance). As we have explicitly demonstrated above, HG with
SCC in reduced space is indeed a consistent theory but quantization of the
resultant non-canonical Dirac Bracket algebra  given in (\ref{hpi})
can be problematic. It would be very convenient if one can construct a gauge
invariant analogue of the HG. Precisely this task will be performed
in this section for the linearized HG. This requires conversion of the mixed
SCC - FCC system to a pure FCC system or more explicitly we wish
to modify the linearized HG with the SCC pair $\chi_1,\chi_2$ to a gauge
invariant theory with one FCC, constructed out of the pair $\chi_1,\chi_2$.
For a generic theory, such as HG having a complicated constraint structure, the
above mentioned task is quite formidable. Fortunately there
is a tailor-made scheme formulated originally by Mitra and Rajaraman
\cite{mitra}
and further developed by Vytheeswaran \cite{vyth}. The idea is to interpret
the original gauge-non-invariant theory (with two SCCs) as a gauge fixed version
of the (to be constructed) gauge-invariant theory with
one FCC. The scheme has been termed as Gauge Unfixing (GU) \cite{mitra, vyth}.
If there are additional FCCs they remain intact on the constraint surface.

The construction proceeds as follows. From a generic pair of SCC
$\chi_1,\chi_2$ choose arbitrarily one of them, say $\chi_1$,
construct the combination,
\be
\chi_1\rightarrow \chi \equiv \{\chi_1, \chi_2 \}^{-1} \chi_1. \label{chifcc}
\ee

This allows one to ignore $\chi_2$ and consider $\chi$ as the single FCC. This
means that one needs to modify the system such that
$\chi$ turns out to be an FCC. A mapping of any generic variable $A$ to its
gauge
invariant extension $A_{GU}$ is given by,
\be
A_{GU} \equiv A - \chi_2 \{\chi, A \} + \frac{1}{2 !} \chi_2^2 \{\chi, \{\chi, A
\} \}
- ....... \label{agu}
\ee

Note that $$ \{\chi, A_{GU} \} = 0$$ by construction. Thus the
extended Hamiltonian system has one FCC $\chi$. One can trivially revert back to
the original system by choosing the gauge $\chi_2$. But
clearly the new system allows much more freedom in choosing any other convenient
gauge.

Let us apply this Gauge Unfixing method to the HG model. We rescale one
of the SCC pair  $\chi_1$ in
(\ref{chi1}) to $\chi$:
\be
\chi(x) \equiv \int d^3 y \{\chi_1(x), \chi_2(y) \}^{-1} \chi_1(y)
= \frac{1}{2} \left(h(x) - \frac{\pa_i \pa_j}{\pa^2} h^{ij}(x)\right).
\label{chigu}
\ee

Using the new FCC $\chi$ (\ref{chigu}), we compute the gauge invariant extension
of (or Gauge Unfixed) Hamiltonian ${\cal{H}}_{GU}$ as
\be
{\cal{H}}_{GU}(x) \equiv {\cal{H}}(x) - \int \pi(y)
\{\chi(y), {\cal{H}}(x) \} d^3 y + \frac{1}{2} \int \pi(y) \pi(z)
\{\chi(z), \{\chi(y), {\cal{H}}(x) \} \} d^3 y d^3 z
$$$$ = {\cal{H}}(x) - \pi^2 (x) + \frac{1}{2} \pi^2 (x) $$$$
= \pi_{ij} \pi^{ij} - \frac{1}{2}\pi^2 - \frac{1}{4} h_{ij}
(\pa^2 h^{ij} - 2 \pa_k \pa^i h^{jk}
+ 2 \pa^i \pa^j h - \delta^{ij} \pa^2 h)
$$$$ - \frac{B}{4} (\pa^k \pa_i h_{jk} - \pa^2 h_{ij}
+ \pa^k \pa_j h_{ik} - \pa_i \pa_j h)
(\pa_l \pa^i h^{jl} - \pa^2 h^{ij}
+ \pa_l \pa^j h^{il} - \pa^i \pa^j h). \label{guhamiltonian}
\ee

It is interesting to note that this Gauge Unfixed Hamiltonian
(\ref{guhamiltonian}) is exactly the ($B$-term dependent) spatial higher
derivative extension of EG Hamiltonian. As we have already remarked, the higher
derivative
$A$ and $C$ terms do not contribute to the linearized theory. This is not
entirely unexpected
since {\it{manifestly}} the symmetry violating parameter $\lambda$ was absent in
the reduced HG model (\ref{hamiltonian}), (\ref{hpi}). This concludes the second
part of our work: constructing a gauge invariant extension of the linearized HG.
Obviously we have not constructed a new theory, (that was not the intention),
but the importance of our work
lies in explicitly showing that there is a systematic way, (Gauge Unfixing
method of \cite{mitra, vyth}), by which one can correctly reproduce a gauge
invariant extension of as involved a field theory as the HG. Our work
strengthens further
the result of \cite{bellorin1, bellorin2} that linearized HG is equivalent to
(higher derivative) EG.

{\it{Discussion}}: Since we have already summarized our results in the beginning
let us conclude by mentioning the work we have undertaken now. It will be truly
interesting if one can carry through this Gauge Unfixing scheme for the full
Horava gravity model because it is bound to generate new non-trivial {\it{gauge
invariant}} extension of the Horava model, containing explicitly the lapse dependent terms. This will surely be another ``improved'' version of Horava gravity.
The possibility of this have already been suggested in \cite{li, bellorin2}. We
hope the present work can act as a stepping stone for this development.

\vskip .4cm

{\it{Acknowledgement}}: It is a pleasure to thank Professor Jorge Bellorin  and Professor William Donnelly for
useful correspondences. Furthermore SG is grateful to Professor Masud Chaichian for discussions on this issue during
his visit to Physics Department, Helsinki University, with an ICTP Regular Associateship grant.

\ed